\documentclass[aps,prb,twocolumn,showpacs,superscriptaddress]{revtex4}

\usepackage{graphicx} 

\begin{document}

\title
{Spin-dependent electronic structure of transition metal atomic
chains adsorbed on single-wall carbon nanotubes}

\affiliation{Department of Physics, Bilkent University, Ankara
06800, Turkey}
\author{E. Durgun}

\author{S. Ciraci}
\email{ciraci@fen.bilkent.edu.tr}

\date{\today}

\begin{abstract}
We present a systematic study of the electronic and magnetic
properties of transition-metal (TM) atomic chains adsorbed on the
zigzag single-wall carbon nanotubes (SWNTs). We considered the
adsorption on the external and internal wall of SWNT and examined
the effect of the TM coverage and geometry  on the binding energy
and the spin polarization at the Fermi level. All those adsorbed
chains studied have ferromagnetic ground state, but only their
specific types and geometries demonstrated high spin polarization
near the Fermi level. Their magnetic moment and binding energy in
the ground state display interesting variation with the number of
$d-$electrons of the TM atom. We also show that specific chains of
transition metal atoms adsorbed on a SWNT can lead to
semiconducting properties for the minority spin-bands, but
semimetallic for the majority spin-bands. Spin-polarization is
maintained even when the underlying SWNT is subjected to high
radial strain. Spin-dependent electronic structure becomes
discretized when TM atoms are adsorbed on finite segments of
SWNTs. Once coupled with non-magnetic metal electrodes, these
magnetic needles or nanomagnets can perform as spin-dependent
resonant tunnelling devices. The electronic and magnetic
properties of these nanomagnets can be engineered depending on the
type and decoration of adsorbed TM atom as well as the size and
symmetry of the tube. Our study is performed by using
first-principles pseudopotential plane wave method within
spin-polarized Density Functional Method.
\end{abstract}

\maketitle

\section{introduction}
Nanotubes interacting with magnetic foreign objects have been a
focus of attention due to the possibility of realizing the
technologically promising area of spintronics in molecular
structures. The ability to produce sizeable changes in the
conductance of a nanotube due to an applied magnetic field has
been one of the driving forces for active research on magnetic
properties of carbon-based structures.\cite{tsuka,ciraci} Due to
their inherent spin asymmetry, the interaction with magnetic
foreign objects, such as adsorbed transition metal (TM)
atoms,\cite{fagan2,fagan3,fagan4} nanoparticles\cite{yang_prl},
and substrates,\cite{ferreira} is likely to cause a spin-dependent
response on the transport properties of the combined structure.
\cite{fagan2,yang_prl,durgun} It is now understood that spin-valve
effect appears to have potential applications in the development
of faster, smaller, and more efficient nanoscale magnetoelectronic
devices.

Costa \emph{et al.}\cite{costa} investigated the indirect magnetic
coupling between two distant magnetic adatoms attached to the wall
of a carbon nanotube. They found that the coupling between TM
atoms is mediated by the electronic carriers and is oscillatory
for metallic armchair  tubes, but monotonic for zigzag nanotubes.
Spin-dependent transport through carbon nanotubes sandwiched
between ferromagnetic electrodes has been studied recently.
Experimental papers dealing with multi-wall carbon nanotubes
(MWNT) have produced results which differ not only quantitatively,
but also qualitatively from one another. For example, reported
maximum GMR values using Co contacts have ranged from
9$\%$\cite{tsuka} to 30$\%$.\cite{zhao}

The interaction of magnetic atoms with nanotubes may result in a
half-metallic (HM) system (namely a metal for one spin direction,
but semiconductor for the opposite spin direction) that is of
interest for spintronic devices\cite{tsuka} as well as
nanomagnets. Since some carbon nanotubes are ballistic
conductors,\cite{tans,frank} the spin polarization induced by
magnetic electrodes (such as Fe, Co, or Ni) can be preserved as
the electrons propagate through the nanotube. To this end, it has
been necessary to know which elements can be best bonded to
nanotubes and how they affect the magnetic properties. Based on
the first-principles Density Functional Theory (DFT) calculation,
Yang \emph{et al.}\cite{yang} found that a Cr or V atomic chain
adsorbed on a metallic armchair carbon nanotube opens up a band
gap for the minority spin states, making the whole system a
100$\%$ spin-polarized conductor. The band gaps of minority spin
bands were 0.49 and 0.44 eV for V and Cr, respectively. The
adsorption of Mn, Fe, Co, or Ni chains led to large but not
complete spin polarization.\cite{yang} Fagan \emph{et al.}
\cite{fagan3,fagan4} studied the structural, electronic and
magnetic properties of Fe chains adsorbed on SWNT. They discussed
several configurations including external and internal geometries
by presenting calculated binding energies, band structures and
magnetic moments. Similarly, Yagi \emph{et al.}\cite{yagi}
investigated the interaction of $3d$ transition metal atoms and
dimers with a single-walled armchair carbon nanotube by
first-principles DFT. They found that Co atoms adsorbed at the
hollow site of internal wall of armchair nanotubes can show
half-metallic behavior.

In this paper we present the spin-dependent properties of TM (Co,
Cr, Fe, Mn, and V) atomic chains adsorbed on the external and
internal walls of zigzag SWNTs. We examined how the spin
polarization varies with radius of SWNT as well as with the type
of TM atoms, which are adsorbed according to well-defined patterns
(decorations). Moreover the strain analysis in radial and axial
directions are performed in order to reveal how robust the
magnetic properties are. Present work is complementary to other
studies, which mainly focused on the metallic armchair nanotubes
and predicted half-metallicity.\cite{yagi,yang,tongay} We note
that while the half-metallicity requiring an integer number of
spin per unit cell can exist only for infinite and ideal systems,
realistic devices can be produced only on finite-size SWNTs, which
are either connected to the metal leads or lie on a substrate. In
this respect, the main issue is to achieve a high
spin-polarization on a finite size SWNT. In order to clarify the
effect of nanotube-size on the spin-dependent electronic structure
we examined also finite systems.

\section{Method}

We have performed first-principles plane wave
calculations\cite{payne,vasp} within DFT\cite{kohn} using
ultra-soft pseudopotentials.\cite{vander} The exchange-correlation
potential has been approximated by generalized gradient
approximation (GGA) using two different functionals,
PW91\cite{gga} and PBE.\cite{pbe} For partial occupancies, we have
used the Methfessel-Paxton smearing method.\cite{methfessel} The
width of smearing has been chosen as 0.1 eV for geometry
relaxations and as 0.01 eV for accurate energy band and density of
state (DOS) calculations. All structures have been treated by
supercell geometry (with lattice parameters $a_{sc}$, $b_{sc}$,
and $c_{sc}$) using the periodic boundary conditions.  A large
spacing ($\sim 10 \AA$) between adjacent nanotubes has been
assured to prevent interactions between them.\cite{vacuum} In
single cell calculations of infinite systems, $c_{sc}$ has been
taken to be equal to the lattice parameter of SWNT and in double
cell calculations, $c_{sc}=2c$. Convergence with respect to the
number of plane waves used in expanding Bloch functions and
\textbf{k}-points used in sampling the Brillouin zone (BZ) have
been tested before analyzing the systems.\cite{convergence} In the
self-consistent potential and total energy calculations the BZ of
nanotubes has been sampled by (1x1x15) and (1x1x11) mesh points in
\textbf{k}-space within Monkhorst-Pack scheme \cite{monk} for
single and double cells, respectively. A plane-wave basis set with
kinetic energy cutoff $\hbar^2 |\textbf{k}+\textbf{G}|^2/2m = 350
eV$ has been used. All atomic positions and lattice parameters
have been optimized by using the conjugate gradient method where
total energy and atomic forces are minimized. The convergence of
calculation is achieved when the difference of the total energies
of last two consecutive step is less than $10^{-5}$ eV and the
maximum force allowed on each atom is less than 0.05 eV/$\AA$. As
for finite structures, supercell has been constructed in order to
yield $\sim 10 \AA$ vacuum space in each direction and BZ is
sampled only at the $\Gamma$-point. The other parameters of the
calculations are kept the same. The binding energy (per atom) of
the adsorbed TM atomic-chain  has been calculated for each
configuration by using the expression,

\begin{equation}\label{equ:binding}
    E_b=\{E_T[SWNT]+E_T[TM-chain] - E_T[SWNT+TM-chain]\}/N
\end{equation}

where $N$ is the number of adsorbed TM atoms per cell. In this
equation, three terms respectively stand for the optimized total
energy of the bare SWNT, TM-chain, and SWNT with adsorbed
TM-chain. All the optimized total energies are calculated in the
same supercell. The spin-polarization at the Fermi level, $E_F$ is
defined as

\begin{equation}\label{equ:polarization}
P(E_F)=[D(E_{F,\uparrow})-D(E_{F,\downarrow})]/[D(E_{F,\uparrow})+D(E_{F,\downarrow})]
\end{equation}

in terms of the density of states of majority and minority spin
states, $D(E_{F,\uparrow})$ and $D(E_{F,\downarrow})$,
respectively. The average binding energy $E_b$, the average
magnetic moment per adsorbed TM atom $\mu$, and $P(E_F)$ have been
calculated for different level of coverage  of  $\theta=$1/2, 1
and 2. Here, $\theta$ indicates the number of adsorbed TM atoms
per unit cell. In order to remove the constraints of supercell
geometry and to test the stability further, the TM atomic
chain-SWNT systems have been relaxed after their supercell sizes
are doubled (namely $c_{sc}=2c$). Moreover, in order to test the
effects of deformation on the physical properties, we have also
studied the cases where the underlying (8,0) tubes are kept under
-25$\%$ radial strain.\cite{oguz1}

\begin{table*}
\begin{center}
\begin{tabular}{c|c|c|c|c|c|c|c|c|c|c|c|c}

~&\multicolumn{4}{c|}{$\theta=1/2$}&\multicolumn{4}{c|}{$\theta=1$}&\multicolumn{4}{c}{$\theta=2$}\\
     \hline\hline%

      ~&$d_{C-TM}(\AA)$&$E_b$(eV)&$\mu(\mu_B)$&$P(E_F)$&$d_{C-TM}$&$E_b$(eV)&$\mu(\mu_B)$&$P(E_F)$&$d_{C-TM}$&$E_b$(eV)&$\mu(\mu_B)$&$P(E_F)$\\
     \hline%
Co&2.0&1.7&1.1&-&2.0&1.4&1.1&-&2.0&0.6&1.7&-0.65\\
Cr&2.2&0.6&4.2&-0.21&2.2&0.4&5.2&0.38&2.3&0.5&4.4&0.53\\
Fe&2.1&0.8&2.2&-0.91&2.1&0.9&4.0&-&2.2&0.5&3.1&-0.65\\
Mn&2.2&0.4&5.5&-&2.2&0.7&5.0&-&2.5&0.6&4.6&-0.19\\
V&2.2&1.4&3.8&0.68&2.2&1.5&4.1&0.90&2.3&1.0&2.9&0.73\\
\end{tabular}
\end{center}
\caption{The distance between TM and nearest C atom $d_{C-TM}$;
average binding energy $E_b$; average magnetic moments per atom
$\mu$; spin-polarization at the Fermi level $P(E_F)$ for chain
structures of Co, Cr, Fe, Mn, and V transition metal atoms
adsorbed on the (8,0) SWNT for $\theta$ = 1/2, 1, and 2. $P(E_F) <
$ corresponds to $D(E_F, \downarrow) > D(E_F, \uparrow)$. }
\label{tab:external}
\end{table*}

\section{TM-wires adsorbed on SWNT}
In this section, we first summarize our results for TM atoms
adsorbed on the external and internal walls of the (8,0) SWNT to
form an atomic chain.

\begin{figure}
\includegraphics[scale=0.5]{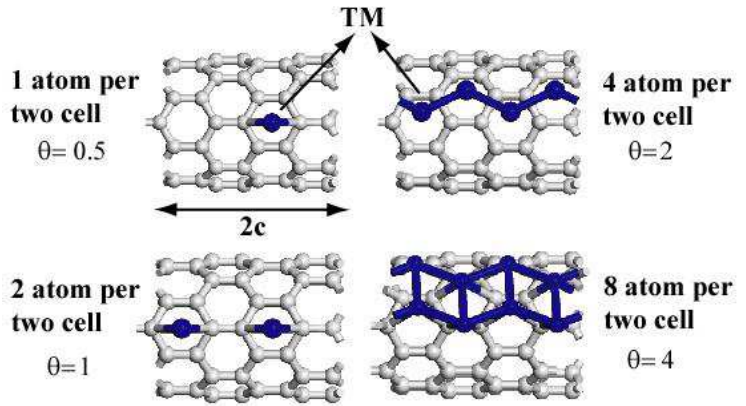} \caption{(Color online)The configuration of adsorbed
transition metal (TM) atoms (Co, Cr, Fe, Mn, and V) forming chain
structures on the (8,0) SWNT are illustrated for various coverage
geometries, such as $\theta$=1/2, 1, 2 and 4.}
\label{fig:positions}
\end{figure}

\begin{figure*}
\includegraphics[scale=0.6]{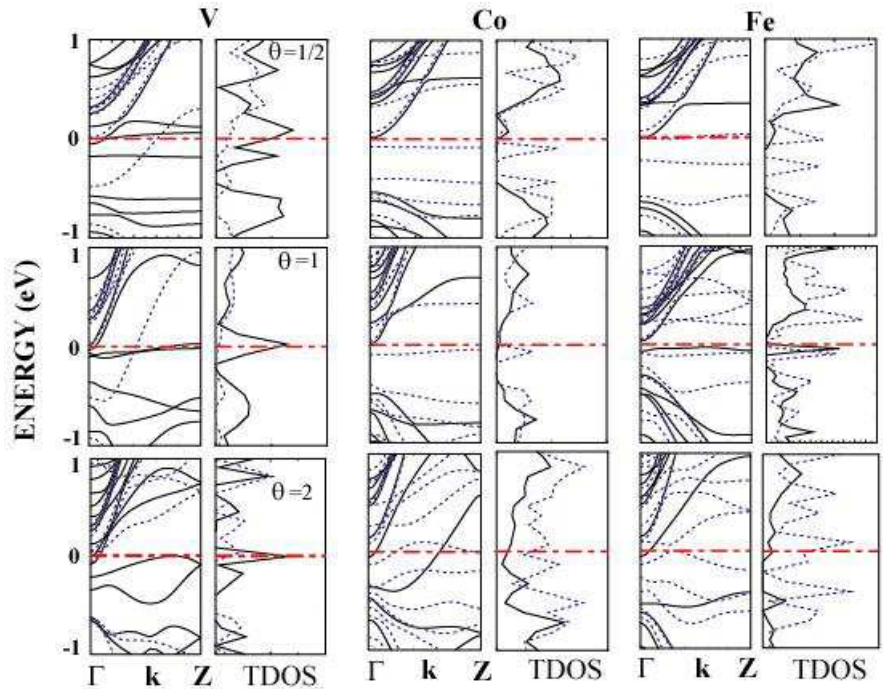}
\caption{(Color online) The spin-dependent band structure and
total density of states (TDOS) of V, Co and Fe-chains adsorbed on
the zigzag (8,0) SWNT for $\theta$=1/2, 1 and 2 geometries. Solid
and dotted lines are for majority and minority spin states,
respectively. The zero of energy is set to the Fermi level $E_F$.}
\label{fig:v}
\end{figure*}

\subsection{External adsorption}
\begin{figure}
\includegraphics[scale=0.45]{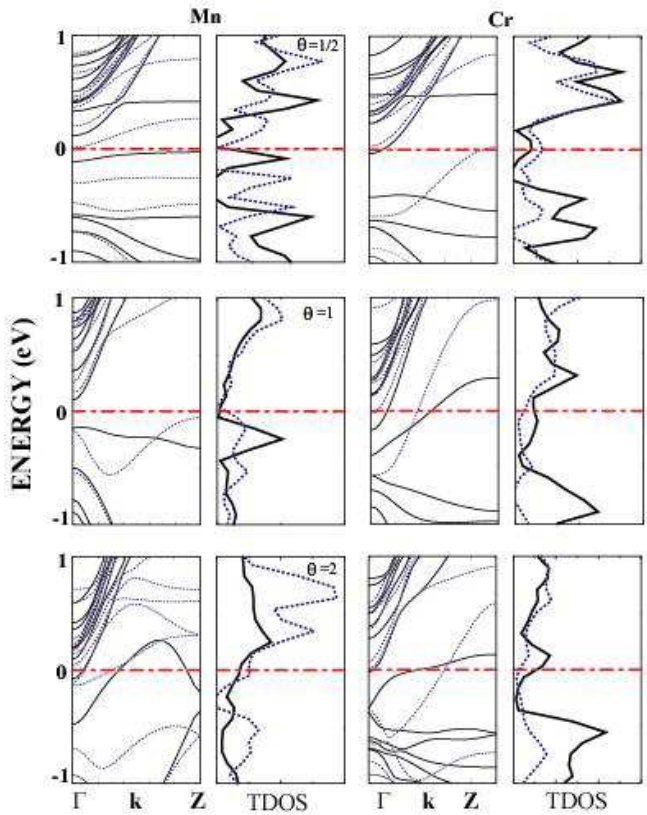}
\caption{(Color online) The spin-dependent band structure and
total density of states (TDOS) of Mn and Cr-chains adsorbed on the
zigzag (8,0) SWNT for $\theta$=1/2, 1 and 2 geometries. Solid and
dotted lines are for majority and minority spin states,
respectively. The zero of energy is set to the Fermi level $E_F$.}
\label{fig:mn}
\end{figure}

Bond distances $d$, $E_b$, $\mu$, and $P(E_F)$ calculated for the
Co, Cr, Fe, Mn, and V atomic chains adsorbed on the (8,0) zigzag
SWNT are listed in Table \ref{tab:external} for $\theta=1/2, 1$,
and $2$. The atomistic model corresponding to various coverages is
illustrated in Fig.\ref{fig:positions}. The spin-polarized band
structures and the total density of states (TDOS) near $E_F$ are
presented in Fig. \ref{fig:v} for adsorbed V, Co, and Fe chains
and in Fig. \ref{fig:mn} for Mn and Cr chains.

\subsubsection{Vanadium}
The ground state of the V-chain adsorbed on SWNT is found to be
ferromagnetic for all geometries described in Fig.
\ref{fig:positions}. The value of the average magnetic moment
\cite{mu}, $\mu$ is calculated as 3.8, 4.1, and 2.9 $\mu_B$ for
$\theta$=1/2, 1, and 2, respectively. The calculations are also
performed for the bare V-chains by removing the (8,0) SWNT, but
keeping the same chain geometry when they were adsorbed on the
tube. For example, for $\theta$=1/2, where the distance between
nearest V atoms ($d_{V-V}$) is 8.52$\AA$, $\mu$ is calculated as
5.0 $\mu_B$, that is equal to the magnetic moment of the free V
atom in $s^1d^4$ configuration. This indicates that for
$\theta$=1/2 V-V coupling is negligible. V-C interaction or charge
transfer between V and C atoms is responsible for the reduction of
$\mu$ from 5.0 $\mu_B$ to 3.8 $\mu_B$ upon adsorption on SWNT. As
for $\theta$=1, $d_{V-V}$ of the bare V atomic chain becomes
4.26$\AA$ and $\mu$=4.8 $\mu_B$; for $\theta$=2, $d_{V-V}$=2.4
$\AA$  and resulting $\mu$ is 4.1 $\mu_B$. These interactions are
also crucial for the stability of decorated structures on SWNT
which were discussed previously both experimentally\cite{zhang}
and theoretically.\cite{durgun} The coupling between V atoms gets
stronger as $\theta$ increases. This causes a slight increase in
the distance between V-C atoms from 2.2 to 2.3 $\AA$.

The adsorption of the V-chain makes the semiconducting (8,0) SWNT
metallic for both spin directions. Complete polarization at E$_F$,
in other words half-metallicity (having an integer number of net
spin in a cell) did not occur. However, for $\theta$=1/2 and 1 the
density of states for majority spin carrier at E$_F$, $D(E_{F},
\uparrow)$ is much larger than minority spin carrier as
illustrated in Fig. \ref{fig:v}. Since, $P(E_{F})$ as large as
$90\%$ can be achieved, these structures may be suitable for
spintronic device applications.

For nanotubes, namely (10,0) and (14,0) with a larger radius, Yang
\emph{et al.}\cite{yang} found $d_{C-V}$ as 2.2 $\AA$ and $\mu =
2.2\mu_B$ with $P(E_F)=45 \%$ for $\theta=2$ geometry. They also
showed that V-chain adsorbed on armchair SWNTs exhibit HM
properties.\cite{yang} Andriotis \emph{et al.}\cite{andriotis}
found that the hollow site of graphene\cite{graphene} is
energetically favorable with $d_{C-V} \sim 1.9 \AA$ and $\mu=1.02
\mu_B$.

\subsubsection{Iron}

For $\theta$=1/2, the SWNT+TM atomic chain system has a
ferromagnetic ground state with $\mu = 2.2 \mu_B$. Here $\mu$ is
reduced from the magnetic moment of free atom due to Fe-C
interaction which, in turn, results in transfer of $4s$ electrons
to $3d$ as confirmed by our Mulliken analysis. The energy band
calculation shows that system is metallic for both types of spin
carriers, but $P(E_F)$ is very high for minority spin carriers.
Analysis of partial density of states (PDOS) suggests that the
hybridized $3d$ states of Fe contributes to $P(E_F)$. For
$\theta=1$, the ground state of the system is still ferromagnetic,
but the increased Fe-Fe interaction and reduction in unit cell
size make the system semiconducting with negligible $P(E_F)$. For
$\theta=2$, the ferromagnetic system is metallic for both spin
carriers. While Fe-C distances have changed slightly under radial
strain $\epsilon_r=$-0.25, the metallicity for both spin carriers
and high $P(E_F)$ is maintained. For the same structure on the
(8,0) SWNT Fagan \emph{et al.}\cite{fagan3} obtained similar
results for ground state properties. They calculated $d_{C-Fe}$
between 2.1-2.4 $\AA$ and $\mu$ as 3.0 $\mu_B$ which are
consistent with present results. However, they obtained $E_b$ as
0.9 eV which is larger than ours.  Our results indicate that the
system is metallic with high $P(E_F)$, whereas they predicted a
semiconducting structure with a small gap. Moreover, the optimized
geometry of the present study is also different. Those differences
between the present study and that of Fagan \emph{et
al.}\cite{fagan3} perhaps originate from different method of
calculations (plane wave versus local basis set). Yang \emph{et
al.} \cite{yang} predicting metallic character with $P(E_F)$=86
$\%$ and $\mu=2.6 \mu_B$ for (14,0), confirms our results. For the
graphene structure Yagi \emph{et al.}\cite{yang} and Duffy
\emph{et al.}\cite{duffy} also found hollow site as the most
stable adsorption site for a single Fe with $\mu \sim 2 \mu_B$.

Finally, we have studied the properties of two parallel Fe-chains
adsorbed on the (8,0) SWNT which is specified as $\theta=4$. The
ground state of the system is ferromagnetic with $\mu=3.0 \mu_B$
per Fe atom and shows metallic behavior for both spin carriers
with negligible $P(E_F)$. Since Fe-Fe interaction is stronger than
Fe-C interaction the Fe atoms show a tendency to form a cluster.
However, for a different geometry, but the same $\theta$, where
two parallel Fe-chains are separated (hence Fe-Fe coupling is
reduced) the net magnetic moment of the ground state did not
change significantly, but $P(E_F)$ increased to $P(E_F)$=0.82.
This result suggests that the spin-polarization is strongly
dependent on $\theta$ as well as on the pattern of the decoration
of the adsorbed TM atoms.

The calculations have been repeated by using different GGA
functional, namely PBE\cite{pbe} for the Fe-atomic chains adsorbed
externally with $\theta$=1/2, 1, and 2. The use of PBE did not
change the results obtained by using PW91.\cite{pbe_result} For
example maximum changes in the binding energy has been 0.1, 0.1
and 0.2 eV for $\theta$=1/2, 1, and 2, respectively.

\subsubsection{Cromium}
Cr atomic chains adsorbed on the (8,0) SWNT with $\theta$=1/2 and
1 give rise to metallic state for both types of spin carriers and
result in negligible $P(E_F)$. However the induced $\mu$ is large.
Bare chains of both $\theta$=1/2 and 1 have the same $\mu=6 \mu_B$
as that of free Cr atom in $s^1d^5$ configuration due to
negligible Cr-Cr interaction. The decrease in $\mu$ when Cr is
adsorbed on SWNT is due to transfer of $s$-electrons into
$d$-electrons. For $\theta=2$ we obtained a significant $P(E_F)$
with $\mu=4.4 \mu_B$. Yagi \emph{et al.}\cite{yagi} reported
$P(E_F)$=29 \% with $\mu = 3.2 \mu_B$ for (14,0) suggesting that
$\mu$ decreases with increase of nanotubes radius. The
calculations by Yagi \emph{et al.}\cite{yagi} on the (6,6) and
(8,8) armchair SWNT showed that a small gap opens for minority
spin carriers and system becomes HM with $100 \%$ spin
polarization with $\mu= 4 \mu_B$. They obtained $E_b$ less than 1
eV for both metallic and zigzag nanotubes which is consistent with
our result indicating relatively  weak interaction between Cr and
SWNT. Duffy \emph{et al.}\cite{duffy} studied single Cr adsorption
on graphene and reported ferromagnetic ground state with $\mu=5.0
\mu_B$


\subsubsection{Cobalt}
While the Co chains adsorbed on the (8,0) SWNT have ferromagnetic
ground states with $\mu \sim$ 1.0 $\mu_B$ for $\theta$=1/2 and 1,
the magnetic moments corresponding to the bare Co chains
($\mu=$3.0 $\mu_B$) are equal to that of single, free Co atom in
the $d^8s^1$ configuration . This indicates that direct Co-Co
interaction is almost negligible at distances larger than 4.26
$\AA$, but Co-C interaction reduces the strength of $\mu$. The
energy band analysis indicates that the Co atomic chains adsorbed
on the (8,0) SWNT for $\theta=1/2$ and 1 is semi-half metallic,
namely the system is semiconducting for minority spin bands, but
the band of majority spin states just touches $E_F$ at the center
of BZ. Moreover, the band originating from localized
$3d_{\downarrow}$ state just below $E_F$ contributes to the
conductance under small bias and makes that spin polarization
significant for minority spin carriers. For $\theta$=2, the system
is metallic for both spin directions with high $P(E_F)$ in favor
of minority spin carrier. The calculations by Yang \emph{et
al.}\cite{yang} indicated also significant spin polarization with
$P(E_F)$= 41$\%$ with $\mu =1.2 \mu_B$ for the Co-chain adsorbed
on the (14,0) SWNT for $\theta=2$. The decrease of $\mu$ with
increasing radius of zigzag nanotubes is consistent with the
results obtained for V, Fe. The calculations concerning the
interaction of Co atom with (4,4) and (8,8) metallic SWNTs at
$\theta$=1/2 indicate that even complete polarization at $E_F$ can
be obtained.\cite{blase}

\subsubsection{Manganese}
The Mn-chain adsorbed on the (8,0) SWNTs has ferromagnetic ground
state for $\theta$=1/2, 1, and 2 geometries. The corresponding
magnetic moments $\mu$ per atom are 5.5, 5.0, and 4.6 $\mu_B$ for
$\theta$=1/2, 1, and 2, respectively. For $\theta$=1/2 geometry,
the magnetic moment of SWNT+Mn chain system is even larger than
that of free Mn atom in $s^1d^6$ configuration. Our PDOS analysis
suggests that Mn-C interaction through the electron transfer from
Mn $4s$ to Mn $3d$ and $4p$ is enhancing the spin
alignment\cite{fagan4}. Even the calculations on the interaction
of single Mn atom with graphene result in a similar charge
transfer from Mn $4s$ to Mn $4p$ and $3d$ orbitals.\cite{duffy}
Bare Mn-chains corresponding to both $\theta$=1/2 and $\theta=$1
have magnetic moments equivalent to that of free Mn atom, since
Mn-Mn interaction is almost negligible for $d_{Mn-Mn}
> 4 \AA$.

The band gap of the bare (8,0) SWNT increases upon the adsorption
of the Mn-chain of $\theta$=1/2. The interesting point is that a
band of majority spin states just touches the $E_F$ at the
$\Gamma$-point exhibiting a half semimetallic character. However,
the system is semiconducting for $\theta=1$, but metallic for
$\theta=2$ with small $P(E_F)$. For $\theta=2$, Fagan \emph{et
al.}\cite{fagan4} predicted similar optimized configuration with
$\mu=4.2 \mu_B$. Yang \emph{et al.}\cite{yang} reported very high
polarization, $P(E_F)$=78 $\%$ with $\mu=3.6 \mu$ for the case of
Mn-chain adsorbed on the (14,0) SWNT according to $\theta=2$.

\begin{figure}
\includegraphics[scale=0.45]{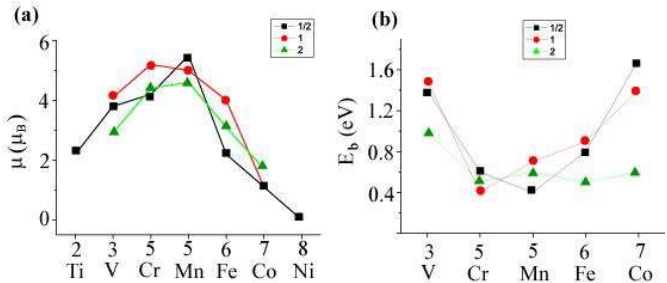}
\caption{The variation of magnetic moment, $\mu$ (a), and the
binding energy, $E_b$ (b) as a function of number of $d$-states
for $\theta$=1/2, 1, and 2 for external adsorption.}
\label{fig:magmom}
\end{figure}

We finally summarize the general trends revealed from the above
discussion. (i) The bond length $d_{C-TM}$ ranges between 2.0
$\AA$ and 2.5 $\AA$; it does not exhibit significant variation
with nanotube radius (R). However, $d_{C-TM}$ slightly increases
with increasing $\theta$ due to the increasing adatom-adatom
coupling. (ii) The binding energy $E_b$, decreases as $R$
increases. This is an expected result due to the curvature effect.
$E_b$ has the lowest value for Mn having half-filled $d$-shell in
$3d^54s^2$ configuration. (iii) Generally $\mu$ decreases as $R$
increases. Maximum of $\mu$ is obtained for Cr and Mn. The
variation of $\mu$ and $E_b$ with respect to the number of
$d$-electrons, $N_d$, of the adsorbed TM atom is plotted in Fig.
\ref{fig:magmom}. Interestingly, different adsorption geometries
corresponding to $\theta$=1/2, 1, and 2 display similar overall
behaviors. The ground state magnetic moment $\mu$ has a maximum
value for $N_d=5$ (corresponding to half-filling of the
$d$-shell). In contrast, $E_b$ shows a minimum at $N_d$=5. Earlier
it has been shown that $E_b(N_d)$ passes through a two maxima for
$N_d$=2 (Ti $3d^2 4s^2$ configuration) and $N_d$=8 (Ni in
$3d^84s^2$ configuration).\cite{durgun} The behavior illustrated
in Fig. \ref{fig:magmom} is explained by using Friedel model [See
Ref \onlinecite{durgun}and \onlinecite{friedel}].

\subsection{Internal adsorption}
The results obtained from the adsorption of Co, Fe, and V-chain
for $\theta$=1/2 and 1 on the internal wall of the (8,0) SWNT are
summarized in Table \ref{tab:internal}. The variation of
$d_{C-TM}$, $E_b$, and $\mu$ with $\theta$ exhibit trends similar
to those in the case of external adsorption. However,
spin-polarization at $E_F$ displays some differences from external
adsorption. For example, while $P(E_F)$ is usually significant for
external adsorption  at $\theta=1/2$, it is negligible for
internal case. Similar to the external counter part, the ground
state of internally doped (8,0) SWNT is ferromagnetic for all
geometries. However, the band structure corresponding to the
internal adsorption usually changes significantly for most of the
cases. This situation shows the effect of confinement on the
interaction between TM and C. Both $\theta=1/2$ and $\theta=1$
geometries of adsorbed V-chains exhibit metallic character, but
the high $P(E_F)$ calculated for the external doping case
diminishes for $\theta$=1/2 and reduces to 0.22 for $\theta$=1.

The change in $\mu$ is more significant when Fe is adsorbed
internally. For $\theta=1/2$, while the Fe-chain externally doped
is metallic with high $P(E_F)$, internally adsorbed system becomes
semiconducting. On the other hand, for $\theta=1$, the
semiconducting system of the external adsorption case shows
metallic behavior with $P(E_F)$=-0.62 for the internal adsorption.
The change in electronic structure as well as in $P(E_F)$ is again
due to the hybridization of $d$-bands. Localized and almost
dispersionless $d$-bands of external chains are dispersed for the
internal case due to increased coupling. The overall shape of the
band structures is similar, but near $E_F$ changes become
significant. For the (4,4) armchair SWNT, Yagi \emph{et
al.}\cite{yagi} also found ferromagnetic ground state with the
same adsorption geometry corresponding to $\theta=1/2$. The atomic
positions and $E_b$ are very close for both case, but just a small
increase in $\mu$ (from 3.0 to 3.1 $\mu_B$) for the internal
adsorption is pointed out.

As for Co, the semi-half metallic system becomes semiconducting
for $\theta=1/2$ and metallic for $\theta=1$ with $P(E_F)$=0.77 in
the case of internal adsorption. The change in the dispersion of
the $d$-band of minority carriers determines the electronic
properties and polarization of the system. The internal adsorption
of Co atoms inside (4,4) and (8,8) armchair SWNT makes the system
half-metallic.\cite{yagi}

Briefly, in the internal adsorption, we see that geometry and
$d_{C-TM}$ do not change significantly with the type of TM atom.
$\mu$ generally decreases for the internal adsorption (except for
Co), since more $4s$ electrons are transferred to $3d$. As the
strength of interaction changes, the value of $E_b$ oscillates and
hinders the derivation of a general rule. Nevertheless, the cases
studied here clearly indicate that the polarization near $E_F$ can
also be manipulated by changing the doping from external to
internal walls of SWNT.

\begin{table}
\begin{center}
\begin{tabular}{c|c|c|c|c|c|c|c|c}

~&\multicolumn{4}{c|}{$\theta=1/2$}&\multicolumn{4}{c}{$\theta=1$}\\
     \hline\hline%

      ~&$d_{C-TM} (\AA)$&$E_b$(eV)&$\mu(\mu_B)$&~$P(E_F)$~&$d_{C-TM}$&$E_b$(eV)&$\mu(\mu_B)$&$P(E_F)$\\
     \hline%
Co&2.0&1.2&1.0&-&2.0&1.6&1.4&0.77\\

Fe&2.2&0.4&2.3&-&2.1&0.4&2.3&-0.62\\

V&2.2&1.5&3.6&-&2.2&1.4&3.8&0.22\\
\end{tabular}
\end{center}
\caption{The distance between TM and nearest neighbor C atom
$d_{C-TM}$; binding energy $E_b$; magnetic moments $\mu$ per TM
atom; polarization at Fermi level $P(E_F$) of various chain
structures of Co, Fe, and V atoms adsorbed inside the (8,0) SWNT
for $\theta$ = 1/2 and $\theta$=1. $P(E_F) < $ corresponds to
$D(E_F, \downarrow) > D(E_F, \uparrow)$. } \label{tab:internal}
\end{table}

\begin{figure*}
\includegraphics[scale=0.7]{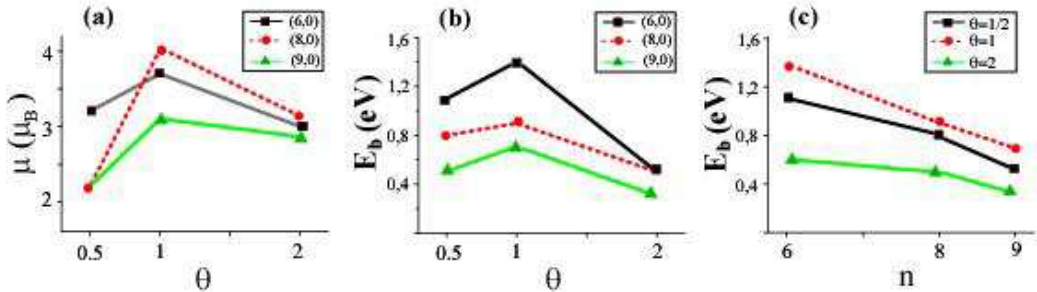}
\caption{Variation of the average magnetic moment $\mu$ and
binding energy $E_b$ of Fe-chain adsorbed on the external walls of
zigzag ($n$,0) SWNT ($n=6, 8, and 9$) with the coverage geometry
$\theta$ and tube index $n$. (a) $\mu$ versus $\theta$; (b) $E_b$
versus $\theta$; (c) $E_b$ versus $n$.} \label{fig:6x0}
\end{figure*}

\subsection{Other type of SWNTs}
In addition to the zigzag (8,0) SWNT, we have analyzed the
interaction of Fe with the (6,0) and (9,0) tubes which are chosen
as prototype for ($n$,0) (where $n$ is the integer multiples of
3). The (6,0) SWNT is metallic due to the dipping of the
$\pi^*$-singlet conduction band into the valence band as a result
of curvature effect\cite{blase}. The (9,0) tube is semiconductor
with very small band gap.\cite{oguz2} All Fe chains ($\theta$=1/2,
1, and 2) have magnetic ground state. Variation of $\mu$ with
$\theta$ is illustrated in Fig. \ref{fig:6x0}(a). $\theta$ as well
as with the index of SWNT ($n$) are shown in Fig. \ref{fig:6x0}(b)
and (c). It appears that $\mu(\theta)$ exhibits similar variation
with $\theta$ for $n$=6,8, and 9. $\mu(\theta=2)$ has comparable
values for all three tubes which have different radii. Owing to
the curvature effect, $E_b$ increases as $n$ decreases (or $R$
decreases). Their electronic band structures display also
interesting properties. For (6,0), the system is metallic for all
$\theta$ and shows high $P(E_F)$ except $\theta=1$. For (9,0), the
minority spin-bands just touch $E_F$ and a small gap opens for
majority spin bands and system becomes almost half-metallic with
perfect spin-polarization at $\theta$=1/2 and 2. For $\theta=1$,
$P(E_F)$ is also high in favor of minority spin carriers but
system becomes metallic for minority and majority spin-bands.

\subsection{Adsorption on finite tubes}

\begin{figure*}
\includegraphics[scale=0.6]{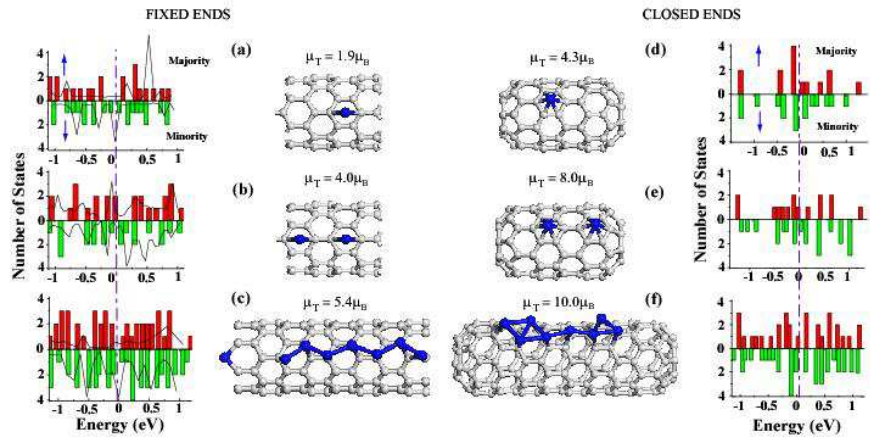}
\caption{Left panels: The number of states and corresponding
structures of Fe atoms adsorbed on a finite (8,0) SWNT with open
but fixed ends for (a) $\theta=1/2$ ( i.e 64 carbon atoms + one
Fe), (b) $\theta=1$ (i.e 64 carbon atoms + 2 Fe), and (c)
$\theta=2$ (i.e 128 carbon atoms + 8 Fe). The zero of energy is
set to the Fermi level $E_F$. The total net magnetic moment for
each finite tube, $\mu$ is shown. Total density of states of
majority and minority spin states of infinite and periodic systems
are also shown by continuous lines for $\theta$=1/2, 1, and 2.
Right panels: Atomic configurations of Fe atoms adsorbed on the
finite size (8,0) SWNT with closed ends. (d) One Fe atom is
adsorbed on a tube consisting of 96 carbon atoms. (e) Two Fe atom
are adsorbed on a tube consisting of 96 carbon atoms. (f) Eight Fe
atoms are adsorbed on a tube of consisting of 160 carbon atoms.
The calculated total magnetic moments $\mu_T$ and TDOS of majority
and minority spin states are illustrated for each configuration.}
\label{fig:finite}
\end{figure*}

While the study of periodic or infinite structures in previous
sections may give an idea about the behavior of the systems in
ideal cases, devices in real applications should have finite size
and may be on substrates and/or connected to the leads. To examine
the finite size effect, we considered  Fe-chain adsorbed on the
finite (8,0) tube. In the first model, we placed Fe atoms
according to $\theta=1/2$ and 1 geometry on a segment of the (8,0)
SWNT consisting of 64 carbon atoms and for $\theta=2$ case on a
segment consisting of 128 carbon atoms. All tubes have open, but
fixed ends. These finite models with fixed ends may be relevant
for SWNTs connected to the electrodes from both ends. In this case
the dangling bonds of free-end carbon atoms are combined with
electrode states. Since this is a finite system all the parameters
of calculation including supercell size are reoptimized as
discussed in Section III. Fe atoms remain stable for $\theta$=1/2
and 1 geometries, but one Fe atom is detached from the wire for
$\theta=2$ geometry and is attached to C atom at fixed ends. For a
finite but longer system this effect will be minute. The ground
state of all the systems are found to be ferromagnetic with total
magnetic moments, $\mu_T$= 1.9, 4.0, and 5.4 $\mu_B$ for
$\theta=$1/2, 1, and 2, respectively. Magnetic moment per Fe
decreased with increased Fe-Fe interaction at $\theta=2$. The
results indicate that ferromagnetic ground state will be conserved
for finite systems. When the number of states near $E_F$ is
compared with TDOS of infinite counter parts (see Fig.
\ref{fig:finite}) we also notice some changes in the distribution
of spin states. These changes occur, since firstly, the electronic
structure of bare nanotube changes due to fixed open ends.
Secondly, the interaction between Fe and C atoms at both ends
affects the electronic structure. Nevertheless, as the length of a
finite-size system increases, the discrete electronic states are
expected to converge to the spin dependent TDOS of infinite and
periodic system. On the other hand, the fact that the contribution
of minority spin states is relatively larger than that of majority
spin states near $E_F$ for $\theta=$ 1/2 and 2 is similar to their
infinite counter parts yielding high $P(E_F)$.

In the right panels of Fig. \ref{fig:finite}, we present more
realistic systems for finite-size devices. Here we consider
slightly longer segments of the (8,0) SWNT and let the carbon
atoms at both ends relax to close. These segments comprises 96
carbon atoms for $\theta=1/2$ and $\theta=1$, but 160 carbon atoms
for $\theta=2$. By adsorbing Fe atoms similar to the cases of
$\theta$=1/2, 1, and 2, we examined geometry and then calculated
spin-polarized electronic structure and magnetic moments. While Fe
atoms remain stable for $\theta$=1/2 and 1, the Fe chain which is
composed of 8 atoms for $\theta=2$ has deformed due to end effects
and strong Fe-Fe interaction. For longer SWNTs this effect is
expected to be minute and in any case the ferromagnetic ground
state is conserved with $\mu_{T}=10 \mu_B$ like the stable low
doping cases $\theta$=1/2 and 1. The energy level diagram of spin
states and total magnetic moments of those Fe adsorbed needles are
strongly dependent on the number of Fe atoms, and on their
adsorption geometry. Moreover, we see dramatic changes between
left panels (corresponding to fixed ends) and right panels
(corresponding to closed ends). The spin-polarization and the
ferromagnetic ground states are expected to be maintained even
after these finite systems are connected to the non-magnetic metal
electrodes from both sides. Depending on the character of the
contact and type of the metal, the discrete levels can shift and
can form resonances. Under an applied electric field these
ferromagnetic needles behave as a resonant tunnelling device, as
well as a spin valve for different spin directions. The size of
the SWNT segment and the geometry of decoration of TM atoms, as
well as their type can be relevant parameters to engineer
nano-spintronic devices.

\section{Conclusion}
This paper presented a systematic analysis for the stability,
atomic, electronic, and magnetic properties of TM atomic chains
adsorbed on the external and internal wall of the (8,0) SWNT. For
the sake of comparison we also considered bare TM chains by
removing SWNT. The effects of coverage, geometry of the adsorbed
chain configuration and the size of the tube on the magnetic and
electronic properties have been investigated. We found that all
adsorbed chains have ferromagnetic ground state. The coupling
among the adsorbed TM atoms and the charge transfer between
adsorbed TM and nearest carbon atom of SWNT play an important role
in determining the resulting magnetic moment. Usually, the
magnetic moment of the free TM atom is reduced upon the
adsorption. We found that high spin polarization at the Fermi
level can be obtained by the adsorption of V- and Fe-chains on the
(8,0) SWNT at specific geometries. The polarization values
achieved as high as 90$\%$ are expected to be suitable for
nanospintronic application. Interesting variation of the magnetic
moment and binding energy with the number of filled $d$-electrons
of the adsorbate have been revealed. The dependence of the
magnetic properties, in particular spin-polarization, on the
radius and band gap of the zigzag tubes are further investigated
by considering TM-chain adsorbed (6,0) and (9,0) SWNTs.

The spin-dependent electronic structure and the net magnetic
moment calculated for finite-size systems are found to be
different from infinite and periodic systems. Our results suggest
that these finite size tubes holding TM atoms can be used as a
nanomagnet and can perform as spin-valve or spin-dependent
resonant-tunnelling devices when they are connected to the metal
electrodes from their both ends. It is demonstrated that the
semiconducting carbon nanotubes constitute a suitable substrate to
hold transition metal chains and metallic leads to form nanoscale
spintronic devices.

\begin{acknowledgments}
SC acknowledges a financial support of T\"{U}BA. We thank Dr. Sefa
Dag for valuable discussions.
\end{acknowledgments}


\begin{thebibliography}{99}
\bibitem{vacuum}In order to ensure a spacing that hinders significant
interaction between nearest SWNTs we took $a_{sc}=b_{sc}=$14, 16,
and 17 $\AA$ for (6,0), (8,0), and (9,0) SWNT, respectively.
\bibitem{convergence} The optimum values of cutoff energy of plane
waves and number of k-points used in the present calculations are
tested for for experimentally well-known crystals such as diamond
for carbon as well as for relevant carbon nanotubes.
\bibitem{mu} All magnetic moments, $\mu$ discussed in the paper
are given per TM atom unless otherwise mentioned.
\bibitem{graphene} It is very well known that SWNT with small
radius, that can be viewed as the graphene sheet rolled over a
cylinder of radius $R$, behaves differently from the planar
graphene (See Ref.\onlinecite{blase}). On the other hand,
adsorption properties of a SWNT with very large radius can be
taken as those of planar graphene corresponding to $R \rightarrow
\infty$.
\bibitem{pbe_result} The geometry and $d_{C-Fe}$ calculated by PW91
are practically same as PBE but, $\mu$ values slightly change and
become 2.4, 4.0, and 3.0 $\mu_B$ for $\theta$=1/2, 1, and 2,
respectively for PBE.
\end{thebibliography}
\end{document}